\documentstyle[preprint,aps,epsfig]{revtex}
\tightenlines
\begin{document}
\draft
\title{
Phase transitions in generalized chiral or Stiefel's models
       }
\author{D. Loison}
\address{
Institut f\"ur Theoretische Physik, Freie Universit\"at Berlin,
Arnimallee 14, 14195 Berlin, Germany\\
Damien.Loison@physik.fu-berlin.de
}
\maketitle
\begin{abstract}
We study the phase transition in generalized
chiral or Stiefel's models using Monte Carlo simulations. 
These models are characterized by a breakdown of symmetry $O(N)/O(N-P)$.
We show that the phase transition is clearly 
first order for $N \ge 3$ when $P=N$ and $P=N-1$, 
contrary to predictions based 
on the Renormalization Group in $4-\epsilon$ expansion but in agreement
with a recent non perturbative  Renormalization Group approach.
\end{abstract}
\vspace{1.cm}
P.A.C.S. numbers: 05.50.+q, 75.10.Hk,05.70.Fh, 64.60.Cn, 75.10.-b 
%\narrowtext

\section{INTRODUCTION}

The critical properties of frustrated spin systems are still under
discussion \cite{Diep2}. In particular no consensus exists about the
nature of the phase transition in canted magnetic systems.
One  example is 
the  stacked triangular lattice with the nearest neighbor antiferromagnetic 
interactions (STA) with vector spins $O(N)$ where $N$ is the number of 
spin components  which is always controversial 
\cite{LoisonSchotteXY,LoisonSchotteHei}. 
The non-collinear ground state due to the frustration leads to a breakdown of 
symmetry (BS) from $O(N)$ in the high temperature to $O(N-2)$ in the low 
temperature. This is different from ferromagnets in which the ground state
is collinear and the BS is $O(N)/O(N-1)$. 
Based on the concept of universality, the class of the 
transition would be different in the two models. 
We generalize this chiral model for a BS of the 
type $O(N)/O(N-P)$. We obtain the STA model for $P=2$  while 
we obtain new BS for $N\ge P \ge 3$. For example, the case $N=P=3$ should 
correspond to real experimental systems, 
this is also applicable in spin glasses where some disorder is present. 
Several authors have already studied
these generalized chiral models applying the Renormalization 
Group  technic \cite{Saul,Zumbach93,Kawamura90}.
In mean field \cite{Zumbach93} for $N> P$ the model shows a usual second 
order type, but for $N=P$ the transition shows a special behavior.
This  last  result could be interpreted with the BS in this case 
being $Z_2 \otimes SO(N)$  and the coupling between the two symmetries 
leading to some special behavior (for example, the case $N=P=2$ in two
dimensions (d) is always very debated \cite{XYfrustre2d}). 
The $d=4-\epsilon$ expansion gives more information. 
The picture is very similar for all $N \ge P \ge 2$ (for details see 
\cite{LoisonSchotteHei} and references therein).
At the lowest order in $\epsilon$, there are up to four fixed points, 
depending on the values of $N$ and $P$. Amongst them are the trivial 
Gaussian fixed point and the standard isotropic $O(NP)$ Heisenberg fixed point.
These two fixed points are unstable. In addition, a pair of new fixed points,
one stable and the other unstable, appear 
if the case is $N\ge N_c(d)$ with 
\cite{Kawamura90} 
\begin{eqnarray}
N_c(d)=5P+2+2\sqrt{6(P+2)(P-1)}-\bigg{[}5P+2
+{25P^2+22P-32 \over 2\sqrt{6(P+2)(P-1)}}\bigg{]}\,\epsilon \,.
\end{eqnarray}
For $P=2$ we find the standard result $N_c=21.8-23.4\,\epsilon$. 
On the other hand, 
for $P=3$ we obtain $N_c=32.5-33.7\,\epsilon$ and 
for $P=4$ we obtain $N_c=42.8-43.9\,\epsilon$.
A "tricritical" line exists which divides a second order region for 
low $d$ and large $N$ from a first order region for large $d$ and small $N$. 
From these results Kawamura, using $\epsilon=1$ ($d=3$), obtained that 
$N_c(d=3)<0$ for all $P$. Thus he concluded that the experimental or 
numerical accessible systems ($N>0$) were in the second order region. 
Unfortunately it has been proved that the results of $4-\epsilon$ are, at best,
asymptotic \cite{LeGuillou}.
They have to be resummed to obtain reliable results.
Indeed for $P=2$ the calculation of the next order in 
$\epsilon$, combined with a resummation technic, leads the experimental
accessible systems for $N=2$ or $N=3$ in the first order region 
\cite{Antonenko 94,Antonenko2}. We  believe that the same applies for 
$P\ge 3$. 
In order to verify our assumption, we have done some simulations 
for $P=3$ and $P=4$
with $N=P$ and $N=P+1$. The most interesting case is $P=N=3$ 
with some possible experimental realizations and connection with 
the spin glasses. Moreover it is meaningful to study the generalized model 
in order to have a better overview. 
The system we analyze is the Stiefel model \cite{Kunz}. This model is
constructed to have the needed BS. It is closely
connected to real systems with complicating interactions, 
which are characterized by the same 
BS (for the case $P=2$ see \cite{LoisonSchotteHei} and
reference therein). From the principle of universality, models with the same 
BS should belong to the same universality class. Moreover we have shown 
that the use of the Stiefel model allows us to avoid problems 
which are seen 
in standard models, such as the presence of a complex fixed point 
(or minimum in the flow) \cite{LoisonSchotteXY,LoisonSchotteHei}.

In the following section II the studied models are presented, we 
describe 
the details of the simulations and the finite size scaling analysis.
Results will be given in section III and the last section is devoted 
to the conclusion.

\section {Stiefel's models, Monte Carlo simulations and first order transitions}
In this section we introduce  different models studied in this work.  

First the $V_{3,3}$ model which is represented in Fig. \ref{figure1}.
The energy of the model is 
\begin{equation}
\label{tata2}
H = J \sum_{ij}  \sum_{k=1}^P 
\Big{[} \ {\bf e}_{k}(i)\cdot{\bf e}_{k}(j) \ \Big{]}
\end{equation}
where the $P$ mutual orthogonal $N$ component unit vectors ${\bf e}_k(i)$
at lattice site $i$ interact with the next $P$ vectors 
at the neighboring sites $j$. The interaction constant is here
negative to favor alignment of the vectors at different sites. 
Taking a strict orthogonality between the vectors is similar to removing
"irrelevant" modes corresponding to the variation between the spins inside
the cell. For example, in the case of a triangular lattice with 
antiferromagnetic interactions (STA) we force the three spins of each cell
to have a rigidity constraint with the sum of all the spins being always zero.
The obtained model is equivalent to the STA at the critical temperature
and can easily be transformed into the Stiefel's $V_{3,2}$ model (for more 
detail see \cite{LoisonSchotteXY,LoisonSchotteHei}).

We did not use the clusters algorithm \cite{LoisonSchotteHei}  
because it gives worse results than the standard Metropolis algorithm
for first order transitions.

The method for choosing the random vector depends on the number of components $N$.
For $N=3$ we follow the method explained in \cite{LoisonSchotteHei} for the 
direct-trihedral model. We construct two orthogonal vectors ${\bf e_1}$ and 
${\bf e_2}$, and the third 
vector is constructed by the vector product of the first two:
\begin{eqnarray}
{\bf e_3} = \sigma \ {\bf e_1} \times {\bf e_2} \ 
\end{eqnarray}
where $\sigma$ is a random Ising variable, corresponding to the Ising symmetry
present in the $V_{3,3}$ model. This is the difference with the 
direct-trihedral model defined in \cite{LoisonSchotteHei}, where no Ising 
symmetry is present. 

To simulate the $V_{4,3}$ model we follow a similar procedure. 
We construct now three orthogonal unit vectors,
${\bf e}_k=(e_k^1,e_k^2,e_k^3,e_k^4)$ with $k=1,\,2,\,3$, 
randomly in four 
dimensions using six Euler angles. 
The first $\theta_0$ must be chosen with probability 
$\sin(\theta_0)^2\, d \theta_0$,
two other with probability $\sin(\theta_{1,2})\, d \theta_{1,2}$
and the rest three $\theta_{3,4,5}$ with probability $\theta_{3,4,5}$.
We obtain for ${\bf e_1}$:
\begin{eqnarray}
e_1^1=&-\cos(\theta_3)*\cos(\theta_1)*\sin(\theta_5)
*\cos(\theta_2)*\cos(\theta_4)&\nonumber \\ \nonumber
&- \cos(\theta_3)*\cos(\theta_1)*\cos(\theta_5)*\sin(\theta_4)\\ \nonumber
& + \cos(\theta_3)*\sin(\theta_1)*\sin(\theta_5)*\sin(\theta_2)*\cos(\theta_0)\\
\nonumber
& + \sin(\theta_3)*\sin(\theta_5)*\cos(\theta_2)*\sin(\theta_4)\\ \nonumber
& - \sin(\theta_3)*\cos(\theta_5)*\cos(\theta_4) \\ \nonumber
e_1^2=& - \sin(\theta_3)*\cos(\theta_1)*\sin(\theta_5)
*\cos(\theta_2)*\cos(\theta_4)\\ \nonumber
& - \sin(\theta_3)*\cos(\theta_1)*\cos(\theta_5)*\sin(\theta_4)\\ \nonumber
& + \sin(\theta_3)*\sin(\theta_1)*\sin(\theta_5)*\sin(\theta_2)*\cos(\theta_0)\\
\nonumber
& - \cos(\theta_3)*\sin(\theta_5)*\cos(\theta_2)*\sin(\theta_4)\\ \nonumber
& + \cos(\theta_3)*\cos(\theta_5)*\cos(\theta_4) \\ \nonumber
e_1^3=& - \sin(\theta_5)*\sin(\theta_2)*\sin(\theta_0) \\ \nonumber
e_1^4=&   \sin(\theta_1)*\sin(\theta_5)*\cos(\theta_2)*\cos(\theta_4)\\
\nonumber
& + \sin(\theta_1)*\cos(\theta_5)*\sin(\theta_4)\\ \nonumber
& + \cos(\theta_1)*\sin(\theta_5)*\sin(\theta_2)*\cos(\theta_0)  \nonumber 
\end{eqnarray}
for ${\bf e_2}$:
\begin{eqnarray}
e_2^1=&\sin(\theta_0)*\sin(\theta_1)*\cos(\theta_3) \nonumber \\ \nonumber
e_2^2=&\sin(\theta_0)*\sin(\theta_1)*\sin(\theta_3)  \\ \nonumber
e_2^3=&\cos(\theta_0)  \\ \nonumber
e_2^4=&\sin(\theta_0)*\cos(\theta_1)  
\end{eqnarray}
and for ${\bf e_3}$:
\begin{eqnarray}
e_3^1=&  \cos(\theta_3)*\sin(\theta_2)*\cos(\theta_4)*\cos(\theta_1)\nonumber 
\\ \nonumber
&+ \cos(\theta_3)*\cos(\theta_2)*\cos(\theta_0)*\sin(\theta_1)\\ \nonumber
&- \sin(\theta_2)*\sin(\theta_4)*\sin(\theta_3) \\ \nonumber
e_3^2=&  \sin(\theta_3)*\sin(\theta_2)*\cos(\theta_4)*\cos(\theta_1)\\ \nonumber
&+ \sin(\theta_3)*\cos(\theta_2)*\cos(\theta_0)*\sin(\theta_1)\\ \nonumber
&+ \sin(\theta_2)*\sin(\theta_4)*\cos(\theta_3) \\ \nonumber
e_3^3=&- \cos(\theta_2)*\sin(\theta_0) \\ \nonumber
e_3^4=&- \sin(\theta_2)*\cos(\theta_4)*\sin(\theta_1)\\ \nonumber
&+ \cos(\theta_2)*\cos(\theta_0)*\cos(\theta_1) \ .
\end{eqnarray}

From this model we can easily create the direct-quadrihedral model
and the $V_{4,4}$, these two models being composed of four orthogonal vectors 
with four components. 
The differences between the two models are the presence 
of an Ising variables in the $V_{4,4}$ where right handed and left handed
are allowed while only one possibility exists in the direct-quadrihedral. 
The direct-quadrihedral and the $V_{4,3}$ models are topologically equivalent,
they should have the same low energy physics and therefore belong to
the same universality class.
The connection between the direct-trihedral 
and the $V_{3,2}$ models \cite{LoisonSchotteHei} is very similar to the above.
We form a fourth
vector from the vector product of ${\bf e_1}$, ${\bf e_2}$ and ${\bf e_3}$:
\begin{eqnarray}
{\bf e_4} = {\bf e_1} \times {\bf e_2} \times {\bf e_3} \
\end{eqnarray}
for the direct-quadrihedral model, and we add a  random Ising 
variable $\sigma$ to the $V_{4,4}$ model:
\begin{eqnarray}
{\bf e_4} = \sigma \ {\bf e_1} \times {\bf e_2} \times {\bf e_3} \ .
\end{eqnarray}

We follow the standard Metropolis algorithm to update one $P$-hedral
after the other.
In each simulation between 20\thinspace 000 to 100 \thinspace 000 
Monte Carlo steps are made for
equilibration and averages. Cubic systems  of linear dimensions
from $L=10$ to $L = 25$ are simulated. 

The order parameter $M$ for this model is
\begin{eqnarray}
M = {1 \over P\,L^3 }\, \sum_{i=1}^P \,\big|\,M_{i}\big|
\end{eqnarray}
where $M_i$ is the total magnetization
given by the sum of the vectors ${\bf e}_i$ over all sites and $L^3$ 
is the total number of sites.

For $N=P$ we define a chirality order parameter:
\begin{eqnarray}
\kappa = {1 \over L^3 }\, {\bf e_P}.(\, \prod_{i=1}^{P-1}\,{\bf e_i}\,)
\end{eqnarray}
where ${\prod}$ means the vector product $\times$.

We use the histogram MC technique
developed by Ferrenberg and Swendsen \cite{Ferren88}
which  
is very useful for identifying a first order transition.

The finite size scaling (FSS) for a first order
transition has been  extensively studied
\cite{Privman,Binder2,Billoire2}.
A first order transition can be identified by some properties and in 
particular by the following:
\begin{itemize}
\item[a.] The histogram $P(E)$ has a double peak.
\item[b.] The magnetization, the  chirality and the energy have hysteresis.
\end{itemize}
The double peak in P(E)
means that at least two states
with different energies coexist in the system at one temperature. 

\section{Results}     

\begin{figure}

\vskip 1cm
\centerline{
\psfig{figure=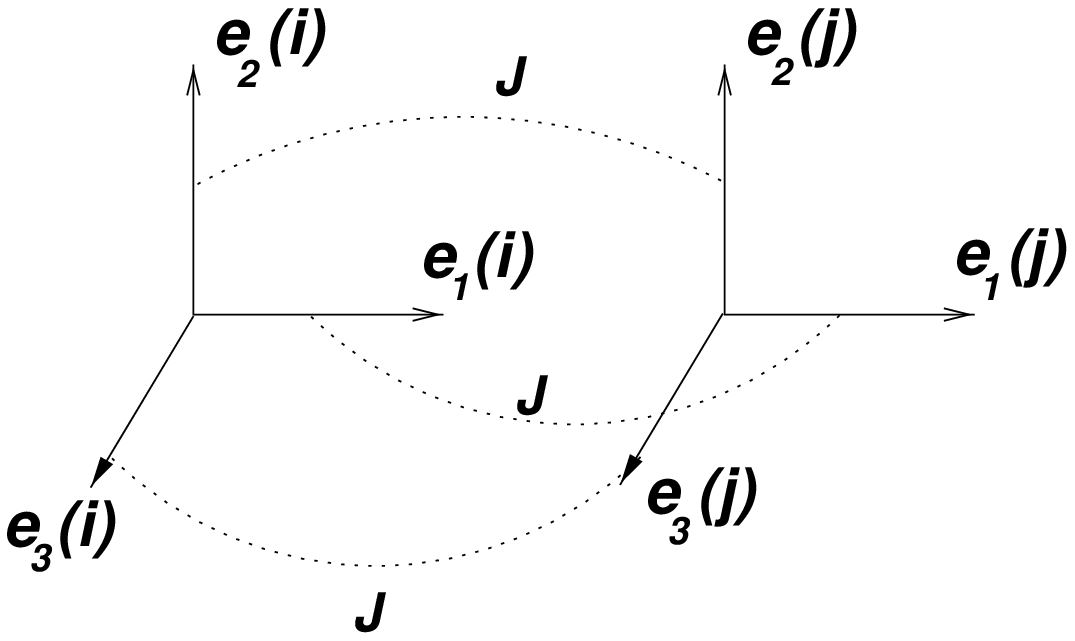,width=6.5cm,height=3.2cm} }
\vskip 1cm
\caption{\label{figure1}
Stiefel's model  $V_{3,3}$
and their interactions. ${\bf e}_1(i)$ interacts only with
${\bf e}_1(j)$, neither with ${\bf e}_2(j)$ which interacts
with ${\bf e}_2(i)$,
nor with ${\bf e}_3(j)$ which interacts with ${\bf e}_3(i)$.
}

\vskip 1cm
\centerline{
\psfig{figure=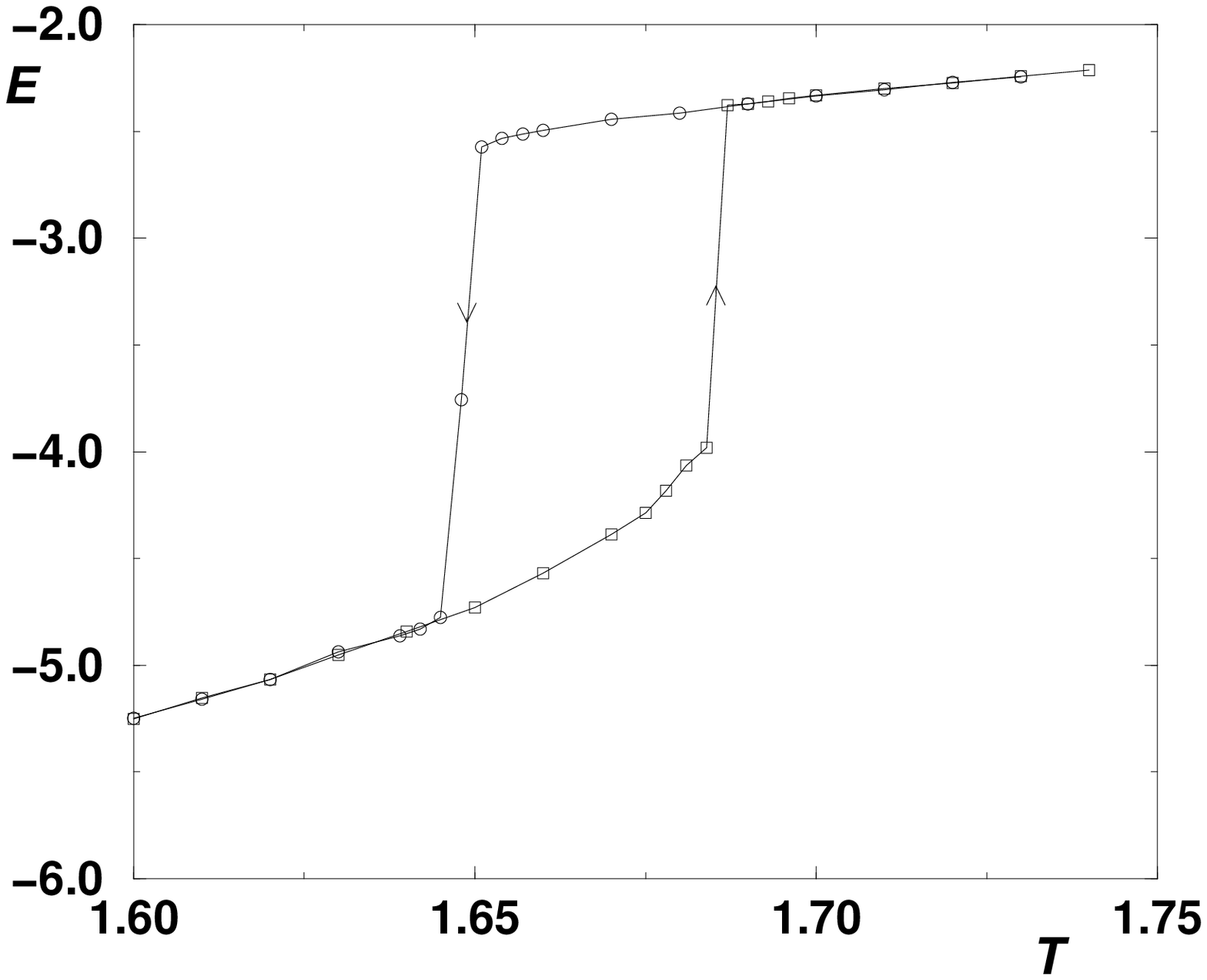,width=6.5cm} }
\vskip 1cm
\caption{\label{figure7}
Internal energy per spin $E$ versus T for the $V_{3,3}$ model.
Lines are guides to the eye. The arrows indicate if the MC simulation
is cooling (circle) or heating (square) the system. The system size
is $L=20$.
A hysteresis is clearly visible.
}

\vskip 1cm
\centerline{
\psfig{figure=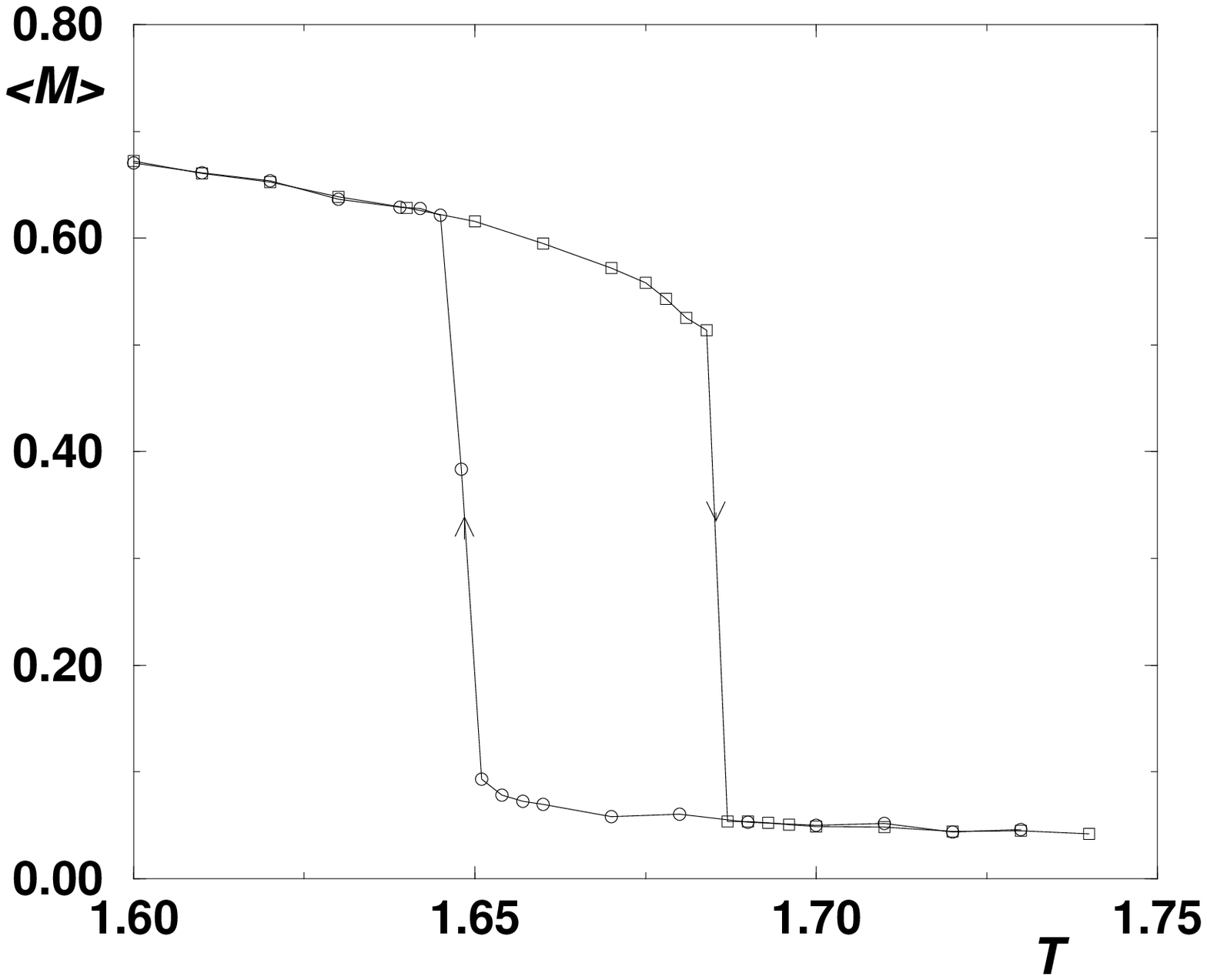,width=6.5cm} }
\vskip 1cm
\caption{\label{figure8}
Magnetization versus T for the $V_{3,3}$ model.
The size of the system is $L=20$.
See comments in Fig.~\ref{figure7}
}

\vskip 1cm
\centerline{
\psfig{figure=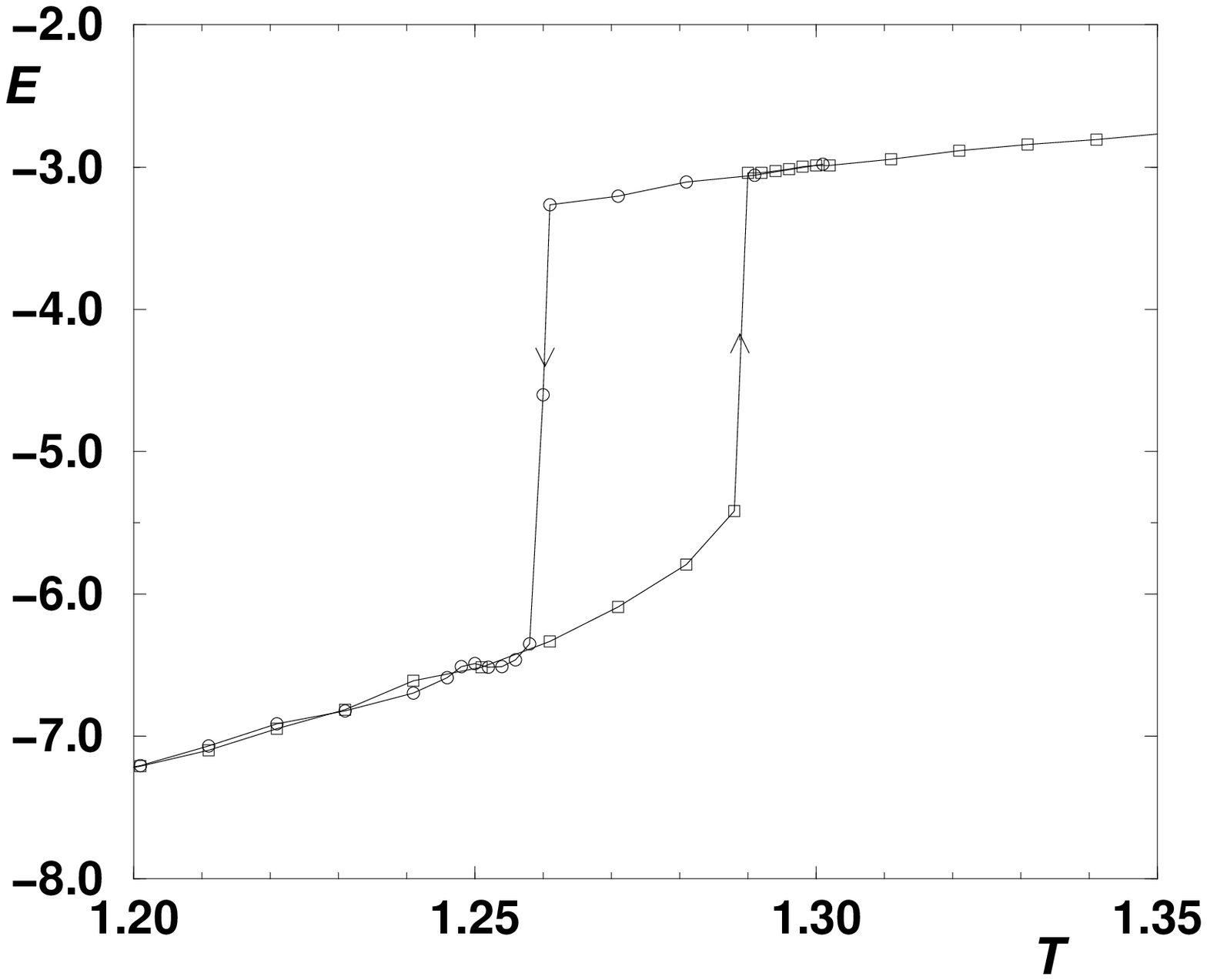,width=6.5cm} }
\vskip 1cm
\caption{\label{figure9}
Internal energy per spin $E$ versus T for the $V_{4,4}$ model.
The system size is $L=10$.
See comments in Fig.~\ref{figure7}
}
\newpage

%\rule{5cm}{0.2mm}\hfill\rule{5cm}{0.2mm}
\vskip 1cm
%\rule{5cm}{0.2mm}\hfill\rule{5cm}{0.2mm}
\centerline{
\psfig{figure=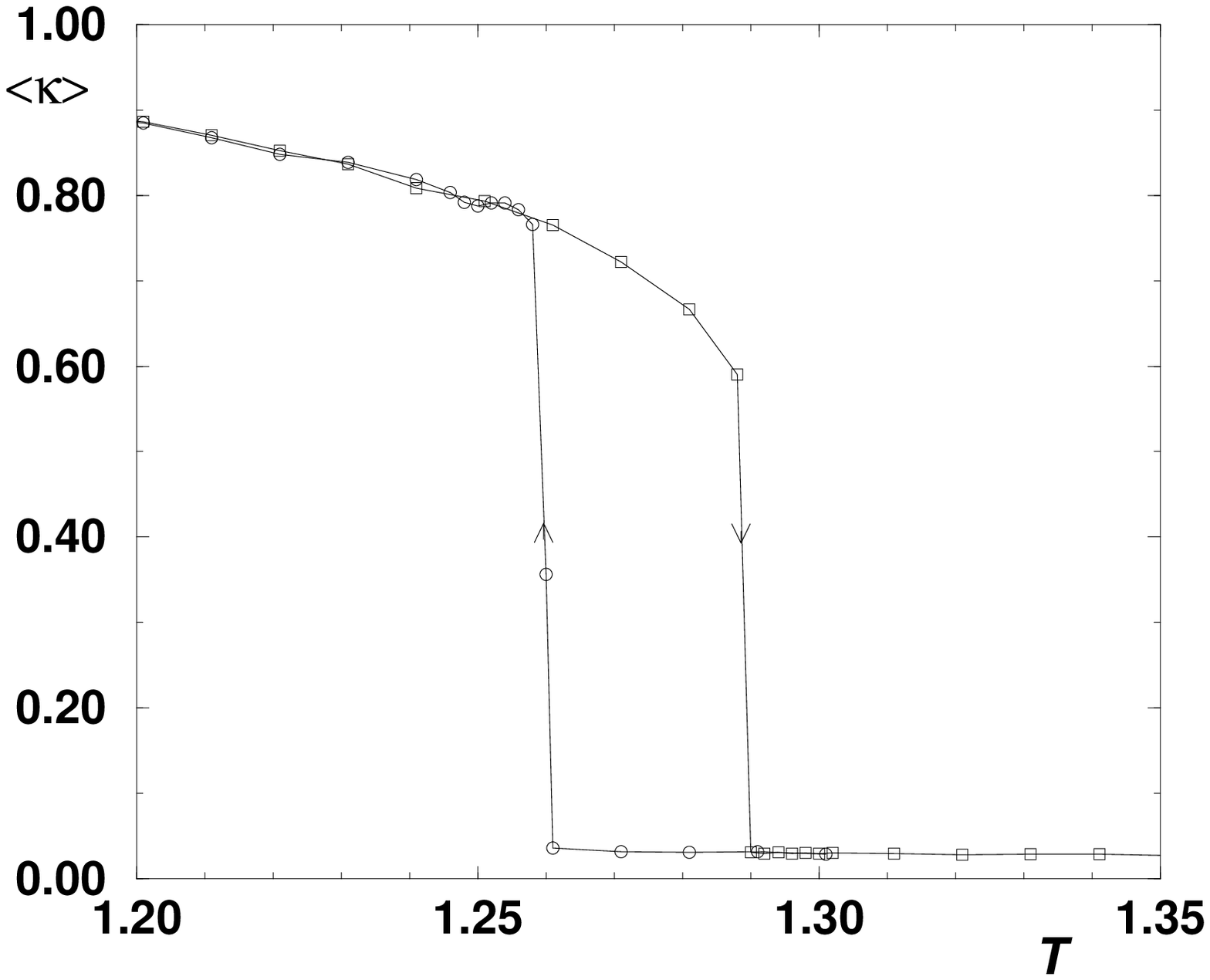,width=6.5cm} }
\vskip 1cm
\caption{\label{figure10}
Chirality versus T for the $V_{4,4}$ model.
The system size is $L=10$.
See comments in Fig.~\ref{figure7}
}

\vskip 1cm
\centerline{
\psfig{figure=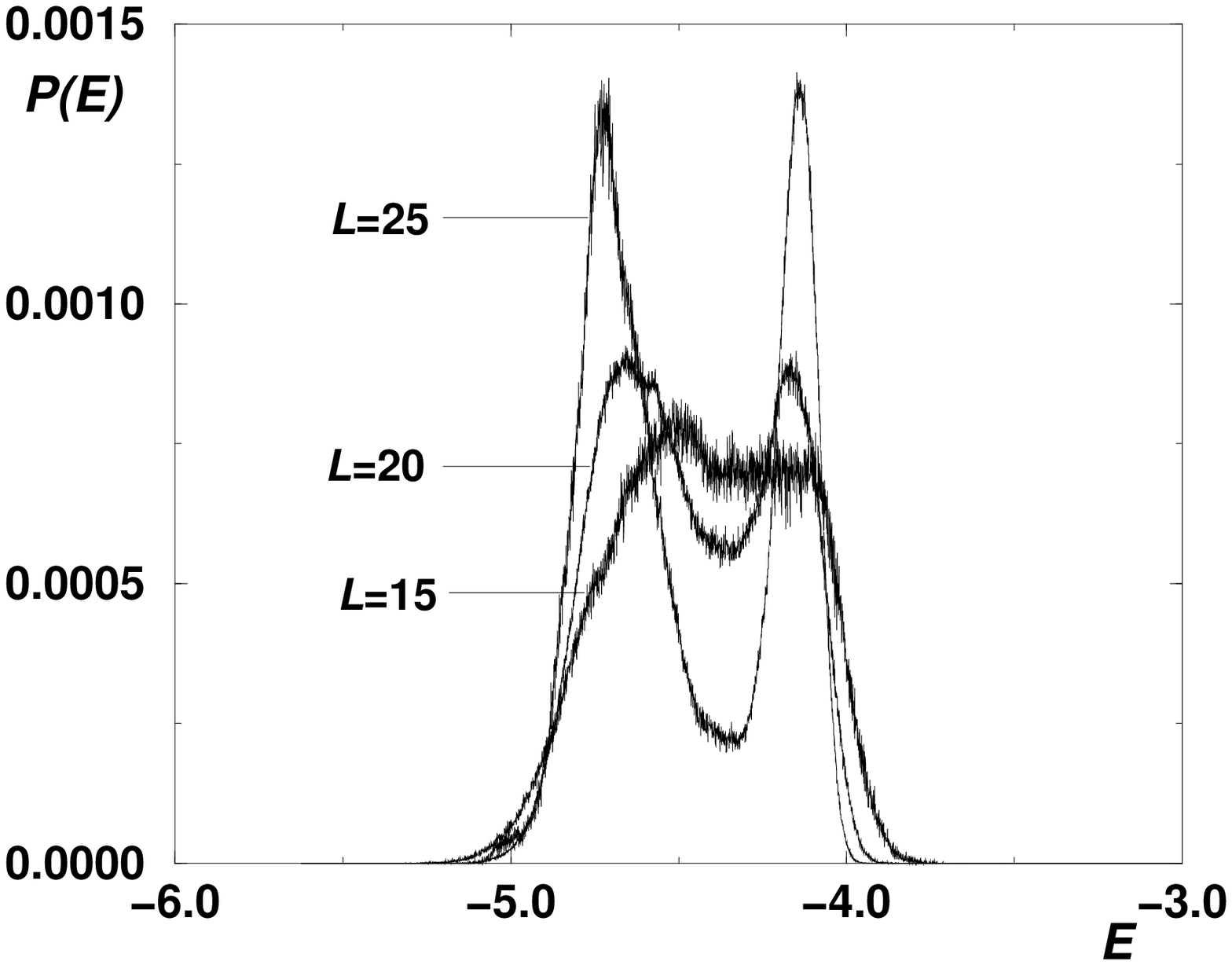,width=6.5cm} }
\vskip 1cm
\caption{\label{figure11}
Energy histogram $P(E)$ as a function of the energy per site $E$ for the
$V_{4,3}$
model for various sizes $L$ at different temperatures of simulation $T_L$:
$T_{15}=1.1802$, $T_{20}=1.1771$,
$T_{25}=1.1758$.
}

%\vspace{2cm}

\vskip 1cm
\centerline{
\psfig{figure=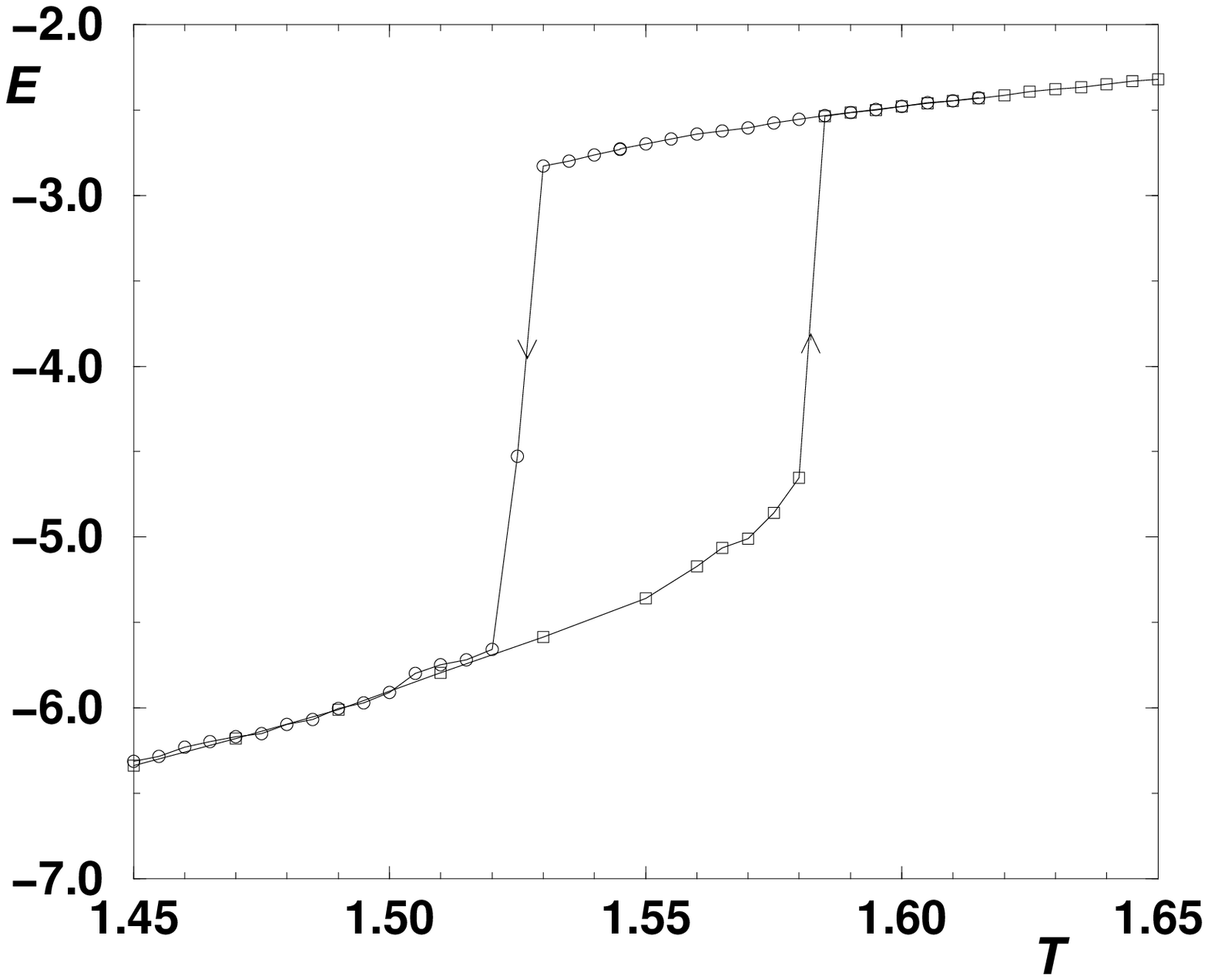,width=6.5cm} }
\vskip 1cm
\caption{\label{figure12}
Internal energy per spin $E$ versus T for the direct-quadrihedral model.
The system size is $L=20$.
See comments in Fig.~\ref{figure7}
}

\vskip 1cm
\centerline{
\psfig{figure=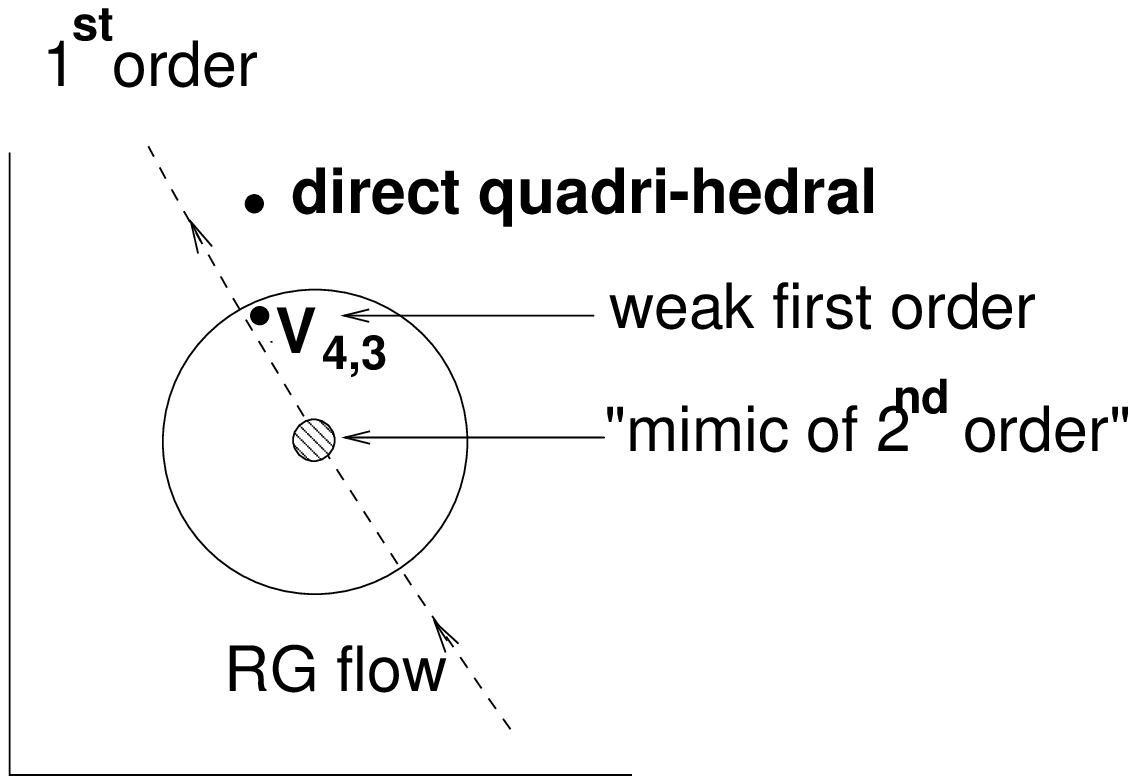,width=6.5cm} }
\vskip 1cm
\caption{\label{figure17}
Hypotheses of Hamiltonian flows induced by
renormalization-group transformations.
The arrows (dashed line) indicate the direction of flow under iteration.
}

\end{figure}

We now present our results for the different models. The $V_{3,3}$
and the $V_{4,4}$
show a strong first order transition. The hysteresis in $E$ 
and $<M>$ are shown in Fig.~\ref{figure7}  and \ref{figure8} for the
$V_{3,3}$ model, and in $E$ and $\kappa$ for the $V_{4,4}$ model in
Fig.~\ref{figure9}  and \ref{figure10}. This is in accordance with the
negative $\eta$ exponent for the $V_{3,3}$ model found in \cite{Kunz} 
which describes a first order transition because $\eta$ must be positive
\cite{LoisonSchotteHei,Patashinskii,Zinn89}.
This result is understandable because there is a coupling between the Ising 
symmetry and the $SO(N)$ symmetry. We notice that the $V_{2,2}$ model 
is also of first order \cite{LoisonSchotteXY} and that the models
have a stronger first order transition if $N$ is greater (we obtain the
same hysteresis for $L=20$ for the $V_{3,3}$ as for $L=10$ for the
$V_{4,4}$). 
Thus we can generalize our
result that the transition is always of first order for $N=P$.

The $V_{4,3}$ model shows no hysteresis. However a double peak structure 
appears in the energy histogram and becomes more apparent when the size
increases (Fig.~\ref{figure11}). 
For greater sizes the two peaks are well
separated by a region of zero probability, the transition time
from one state to the other grows exponentially with the size of
the lattice.
We should obtains hysteresis in the thermodynamic quantities
when the simulation is not too long. 
The $V_{4,3}$ model
has a first order transition but weaker than the direct-quadrihedral
model which, for similar sizes, shows hysteresis.
As explained above the two models belong to the same  
universality class, i.e. a first order transition, 
similar to the  dihedral model $V_{3,2}$ and the
direct-trihedral model \cite{LoisonSchotteHei}.
The addition of the fourth leg to the $V_{4,3}$ model allows 
the first order behavior to be more clearly visible. 
In Fig.~\ref{figure17} we have plotted our hypothesis for 
the RG diagram flow.
Following the initial point, the flow could be under 
the influence of a "complex" fixed point (or minimum of the flow
\cite{Zumbach93}) and the system  mimics a second order transition. 
Well outside the influence of this fixed point the transition is strongly
of first order and in the crossover between these two regions the transition
is weakly first order. For a more developed discussion see 
\cite{LoisonSchotteHei}. 

\section {Conclusion}

We have tried to give a general picture of the transition with a 
$O(N)/O(N-P)$ breakdown of symmetry. We have shown by numerical 
simulations that for 
$N=P=3$ and $N=P=4$, the transition is clearly of first order.   
We have generalized our result for all $N=P$.
A similar conclusion is obtained for $N=4$ and $P=3$. Using the fact 
that for $N=3$ and $P=2$ the transition is also of first order,
we can generalize our result for all $P=N-1$. 
This is in contradiction with 
to the conclusion of Kawamura \cite{Kawamura90} which is based on two loops 
of a $4-\epsilon$ expansion. As we have noted the $\epsilon$
expansion has to be resummed to obtain reliable results. We can try
to achieve this by forming simple Pad\'e approximants. For a function 
$f=a+b\,\epsilon$ we obtain the approximation $f=1/(1-\epsilon \, b/a)$
which we apply to $\epsilon=1$ ($d=3$) and $P=$2, 3 and 4. We obtain 
$N_c(P=2)\sim 10$, $N_c(P=3)\sim 16$ and $N_c(P=2)\sim 21$. 
Unfortunately the results can not be perfect and in 
particular the result for $P=2$ is not close enough to the result 
including the next order expansion $N_c=3.39$ \cite{Antonenko2}
which demonstrates that the lower-order $\epsilon$ expansions are useless 
in this case. However we remark that the $N_c$ "resummed" increases
with $P$ which is in agreement with our result, i.e. that the initial
point in the renormalization flow is farther away from the mimic of the
second order region \cite{LoisonSchotteHei}. Thus the systems will show a 
stronger first order transition. 
This result matches with a recent study on the case $P=N=3$ 
which is based on a non perturbative Renormalization Group procedure
\cite{Delamotte99}.
We conclude that transitions for $N=P$ and $N=P+1$ are of first order 
for all $N$.

\section {Acknowledgments}
This work is supported by the Alexander von Humboldt Foundation.
The authors are grateful to Professors B. Delamotte, G. Zumbach,  and 
K.D. Schotte for discussions.

\newpage
\twocolumn

\end{document}